\documentclass[twocolumn,showpacs,preprintnumbers,amsmath,amssymb]{revtex4}
\usepackage{graphicx}
\usepackage{dcolumn}
\usepackage{bm}
\usepackage{amssymb}
\usepackage{amsmath}
\usepackage{fancyhdr}
\begin{document}

\title{Point-Contact spectroscopy on electron-doped Sr$_{0.88}$La$_{0.12}$CuO$_2$ thin films}

\author{L. Fruchter, F. Bouquet and Z.Z. Li}%
\affiliation{Laboratoire de Physique des Solides, Univ. Paris-Sud, CNRS, UMR 8502, F-91405 Orsay Cedex, France}
\date{Received: date / Revised version: date}

\begin{abstract}{We report on the spectra of point-contacts made on Sr$_{0.88}$La$_{0.12}$CuO$_2$ thin films. Besides a clear evidence for the superconducting gap, we discuss the origin of specific features, such as resistance peaks at the gap voltage and the occurrence of a two-steps resistance decrease.}
\end{abstract}


\maketitle

The so-called infinite-layer compound distinguishes itself from the other members of the cuprate superconductor family by the simplest crystallographic structure, which consists of a stack of CuO$_2$ planes, intercalated by rare-earth planes. The infinite-layer compound may actually be prepared both as a hole-doped material, (Sr$_{1-x}$Ca$_x$)$_{1-y}$CuO$_2$ \cite{Hiroi1991,Azuma1992}, or an electron-doped one, Sr$_{1-x}$La$_{x}$CuO$_2$ (SLCO) as in the present study. These materials are however only obtained under high-pressure synthesis and no single crystals could be grown this way. Growing thin epitaxial film allows one to circumvent this difficulty. In addition, it is possible in this case to tune doping, by modifying the excess apical oxygen content\cite{Li2009}. The symmetry of the order parameter in the electron-doped material -- conventional or with nodes -- has been the subject of several studies, yielding conflicting results\cite{Chen2002,Liu2005,Liu2007, Khasanov2008,White2008,Tomaschko2011,Fruchter2010}. Recently, phase-sensitive measurements on artificial junctions in SLCO thin films found evidence for d-wave pairing\cite{Tomashko2012}. However, the proposal that some electron-doped materials may exhibit an order parameter with both a conventional character and an unconventional one, depending on the contributions of the various parts of a complex Fermi surface\cite{Das2007}, certainly needs further studies. Here, we investigate point-contact spectra made on SLCO films. Besides a superconducting gap value in line with previous determinations, we find that contacts with a low $T_c$ exhibit a two step resistance decrease. The possibility that it could be a genuine spectral feature is discussed.

\begin{figure}
\includegraphics[width= \columnwidth]{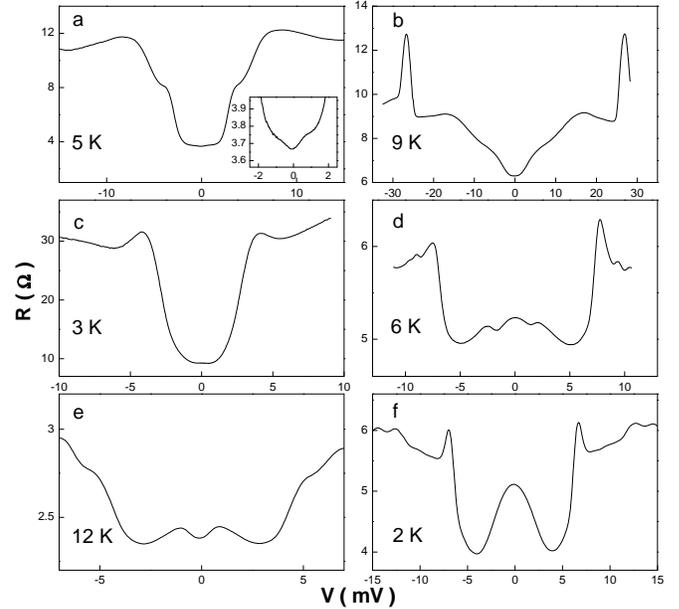}
\caption{Point-contact spectra obtained for several contacts and samples. For the contacts a,b,d $T_c \simeq$ 13 K, and 22 K for e,f.}\label{misc}
\end{figure}

We used 50 nm ~thick films grown by rf-magnetron sputtering on a (100) KTaO$_3$ substrate\cite{Li2009}. These films grow epitaxially with the (001) direction normal to the film. After growing the epitaxial film, an SLCO amorphous layer, about 10 nm thick, was deposited \textit{in situ} at room temperature. Such layer was found to protect the films from degradation, which is otherwise observed after a few weeks in air. It also provides an electrical insulating layer above the film which we used to elaborate some contacts. Different doping where obtained using different annealing time of the samples in vacuum, which varies the oxygen content\cite{Li2009}. Making contacts on cuprates suitable for point-contact studies is notoriously difficult (for a review, see Ref.~\onlinecite{pcs2005}), in particular when the material cannot be cleaved easily, as is the case for the present material.  Contacts were made here in two different ways. In the first method, following the deposition of the protection layer, a gold layer was evaporated on the film. The gold layer was further patterned, leaving a checkerboard of dots about 250 $\mu$m large. The impedance of these contacts was initially rather large,  but it could be tuned down to a few ohms, by applying the tip of an ultrasonic bonding machine on the dots prior to the measurement, or by pressing a tungsten tip on the gold contact and vibrating it \textit{in situ}. These preparations likely induce holes through the amorphous layer and provide contacts with a lower resistance than the original one. Although the resistance of such contacts could be found to increase upon aging at room temperature or thermal cycling, they were stable at low temperature and allowed for the study of the contacts through the whole superconducting range. In the second method of making contacts, the substrate and the film were cleaved and quickly introduced in the measurement chamber, filled with helium gas at a reduced pressure, and cooled down. A low-impedance contact was established by pressing a thin gold strip perpendicular to the broken edge. The differential resistance was measured using standard lock-in techniques, using an AC excitation current, in order to avoid measuring the macroscopic contact non-linearity, and a point-contact voltage inducing a broadening well below the thermal one, $V_{ac} \ll\,$k$_BT/e$.

We investigated several films, each of them with several contacts (either by the first or the second method). Although these contacts displayed a wide range of behaviors, which may also depend on temperature or magnetic field, these observations allow to establish generic features for the spectra, all displayed in Fig.~\ref{misc}. First, many contacts (although sometimes restricted to some temperature or field range) exhibit a two-step decrease of the resistance (Fig.~\ref{misc}a,b,d). The low voltage resistance drop may exhibit a broad maximum at zero-bias (Fig.~\ref{misc}d-f). It may also show a narrow resistance drop at zero-bias (Fig.~\ref{misc}a,e). Although this narrow drop appears to be generally concealed by the broad maximum, both features may sometimes be observed on the same spectrum (Fig.~\ref{misc}e). A feature similar to the broad maximum at zero-bias in the low voltage drop may sometimes be observed for the large voltage drop also (Fig.~\ref{misc}d). Finally, either drops may exhibit a peak or a hump at the edge of the drop (Fig.~\ref{misc}a-d,f). Understanding these various behaviors is clearly a challenge. Indeed, besides non-spectral features, the point-contact spectrum is determined by several unknown parameters, which may be extrinsic: spectral broadening by disorder, interface reflectivity and interface orientation (or even faceting) or intrinsic: superconducting gap symmetry and the material Fermi surface.

\begin{figure}
\includegraphics[width= \columnwidth]{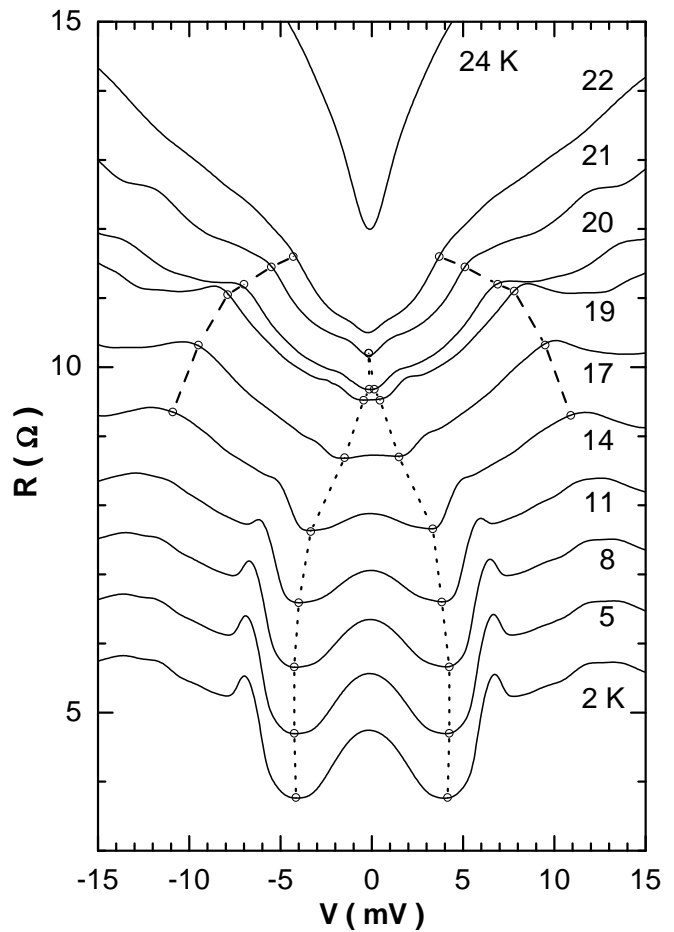}
\caption{Spectra for a sample with bulk $T_c$ = 25 K (spectra have been shifted 0.3$\Omega$/K). The dotted line marks an estimate for a superconducting gap, obtained from the minimum in $R(T)$. The dashed line marks the onset of some resistance drop at larger energy.}\label{temp}
\end{figure}

The double-minimum structure as in Fig.\ref{misc}f is considered as the hallmark of the Andreev-reflection process, as brought into play in BTK model\cite{Blonder1982}. Indeed, at $T = 0$, in the absence of a barrier at the N-S interface, incoming electrons within the superconducting energy gap, $\Delta$, of the Fermi energy experience Andreev reflection, hence a decrease by a factor 2 of the electrical resistance, yielding a 'U-shaped' spectrum (for the simplest case of a conventional s-wave superconducting gap). The effect of a barrier at the interface - due to either elastic scattering or to material Fermi velocity mismatch - is to reflect part of the incoming electrons and thus to limit the Andreev conversion process, yielding the broad resistance increase at zero-bias. Finite temperature and broadening of the quasi-particle density of states both reduce the resistance drop and broaden the spectrum. A reliable estimate of the superconducting gap value may obtained within BTK model from the minimum in $R(V)$, as long as $\Delta \gtrapprox k_B T$. A similar analysis was found to be able to provide an estimate for the superconducting gap for YBa$_2$Cu$_3$O$_{7-\delta}$ in Ref.~\onlinecite{Goll1992}.

\begin{figure}
\includegraphics[width= \columnwidth]{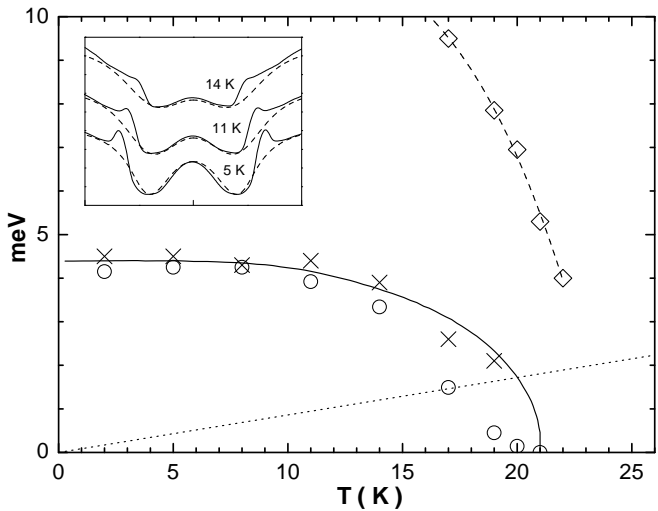}
\caption{Characteristic energies obtained from the data in Fig.~\ref{temp}. Circle is the superconducting gap obtained from the resistance minimum. Cross is the superconducting gap obtained from a fit of the double minimum with BTK model (dashed line in inset, $Z \simeq 0.45$, $\Gamma \simeq 0.07$). Diamond is the onset for the resistance depression at large energy. The dotted line is $V = k_BT/e$; the full one is BCS gap, using $\Delta(0)/k_BT_c$ = 2.4, and the dashed one is a guide to the eye.}\label{gap}
\end{figure}

Fig.~\ref{temp} displays the evolution with temperature of a spectrum with a clear double minimum structure at low temperature. As may be directly read from the spectra, a superconducting gap -- $\Delta\approx 4.5$ meV at $T = 0$ -- closes at $T$= 20 - 22 K, close to the bulk $T_c$ = 25 K for this sample. This value is in line with the one of 13 meV for samples with $T_c$ = 40 K\cite{Chen2002,Khasanov2008}. A fit using BTK model\cite{Plecenik1994} is unable to account for the peaks observed at voltage larger than $\Delta/e$. However, a fit of the double minimum (Fig.~\ref{gap}, inset) allows to determine a superconducting gap value, which is somewhat larger than the direct estimate as $\Delta \simeq k_B T_c$, as expected .

At high temperature, the double minimum vanishes and is replaced by a resistance drop (see e.g. Fig.~\ref{temp}, $T$ = 20 K). At the same time, a drop at larger energy may be observed. Although it is difficult to assign a characteristic energy to this drop, the presence of hump at its edge (as is also the case for the double minimum structure) provides some characteristic energy, which is plotted in Fig.~\ref{gap}, as well as the smaller gap.

\begin{figure}
\includegraphics[width= \columnwidth]{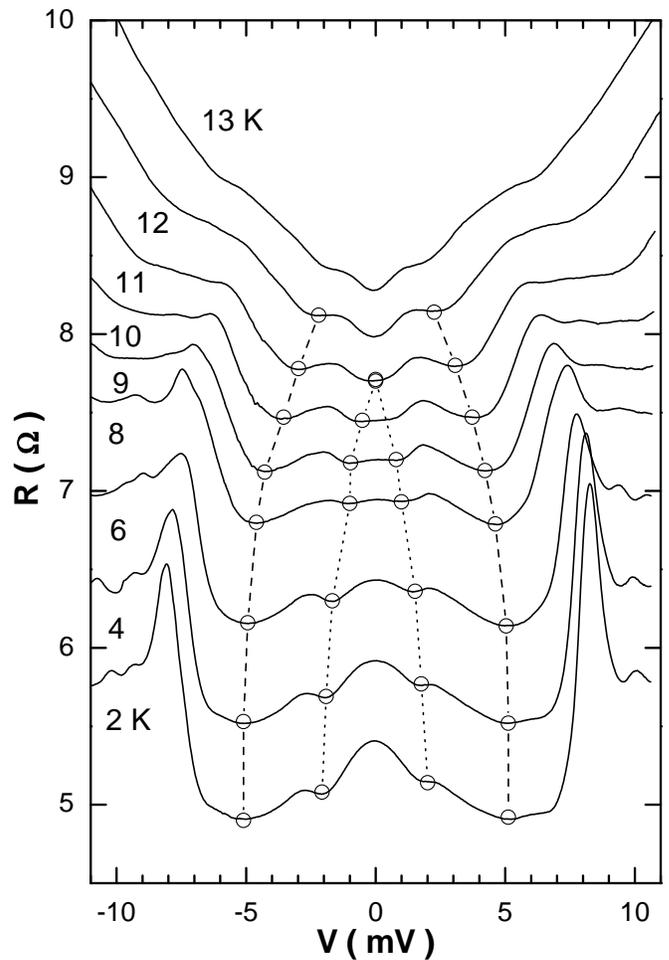}
\caption{Spectra for a sample with bulk $T_c$ = 20 K (spectra have been shifted 0.3$\Omega$/K). The dotted lines mark an estimate for two superconducting gaps, obtained from the minimum in $R(T)$.}\label{temp2}
\end{figure}

\begin{figure}
\includegraphics[width= \columnwidth]{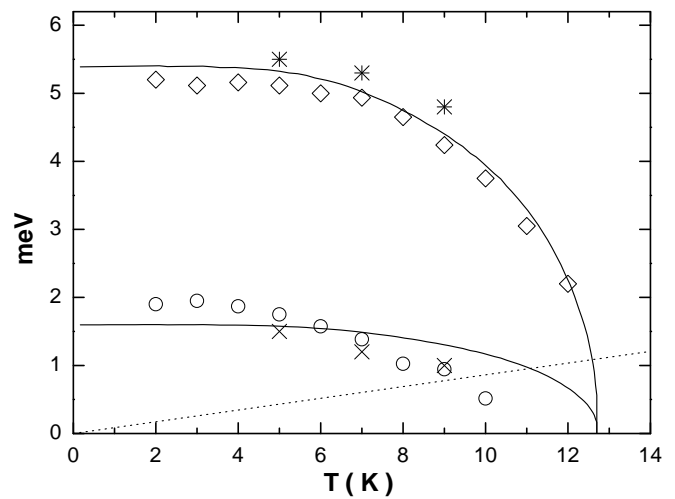}
\caption{Gaps as obtained in Fig.~\ref{temp2}. Crosses and stars are from a fit of the double minimum structure, using two gaps. The dotted line is $V = k_BT/e$.}\label{gap2}
\end{figure}

While the high energy feature in Fig.~\ref{temp} is only visible close to the transition temperature for the contact, for some others two contributions to the contact resistance may be tracked over the whole temperature range. As may be seen in Fig.~\ref{temp2}, the double minimum generally associated to the Andreev-reflection may also be observed for both resistance drops. This suggests that each drop might be associated to a superconducting gap. These gaps, obtained either directly from the resistance minimum or a fit of the double minimum (again, the BTK model is unable to account for the peaks at the edge of the structure) are plotted in Fig.~\ref{gap2}. It is difficult to determine whether both gaps close at exactly at the same temperature, from the data in Fig.~\ref{gap2}, as the smaller gap is comparable to $k_B T$ in a large temperature interval, which invalidates a simple estimate of this gap from the minimum in $R(V)$.

To discuss whether these characteristic energies may have a non spectral origin, we believe it is necessary to gain a better understanding of this peculiar peaked double-minimum structure. Similar spectra, where Andreev-Reflection appears attested by the presence of the double resistance minimum, while $dV/dI(V)$ at the edge of the structure presents humps or peaks and is too steep to be fitted by the BTK model, have already been observed for other cuprates\cite{Yanson1988,Goll1992,Akimenko1992}. The humps - which can be quite sharp (see Figs.~\ref{temp},~\ref{temp2}) cannot be due to multiple Andreev reflections (in case the tip creates a SS contact), as one would then expect sub-gap occurrences at $V=2\Delta/n$, where $n =$ 1, 2, 3..., that we did not observe. Another model assumes that the proximity effect induces superconductivity in the normal electrode. This induces a hump in a voltage range delimited by the superconductor's gap and some average gap of the normal metal layer affected by the proximity effect\cite{Strijkers2001}. However, as underlined in Ref.~\onlinecite{Strijkers2001}, the hump displays a typical width $\Delta/2$, presumably because some average gap $\sim$ $\Delta/2$ is at play in the proximity layer. The resistance hump that we observe is very often much narrower than this. 

One may also expect some contribution from a diffusive contribution, in addition to the ballistic one, yielding non spectral features. Indeed, assuming a purely ballistic regime, one may evaluate the contact dimension in the following way, using for the contact resistance in the normal state (that is, for $eV \gg \Delta$) the Sharvin expression :

\begin{equation}
R_{sh} = \frac{4\pi^3}{S_c S_F} (\frac{e^2}{\hbar})
\label{kulik}
\end{equation}

where $S_c= \pi a^2$ is the contact surface and $S_F= \pi k_F^2$ is the extremal cross-section of the Fermi surface\cite{Kulik1977}. In Eq.~\ref{kulik}, $S_F$ can be approximated by the total Fermi surface. Using approximately 50\% of the first Brillouin zone, one obtains a typical value $a \simeq 400 (R_S\left[\Omega\right])^{-1/2}$ \AA; for a contact with a typical resistance 10 $\Omega$, $a$ is found as large as 130 \AA. On the other hand, using the Sommerfeld expression for the mean free path, $l$(\AA)$= 92\,(r_s/a_0)^2 / \rho$, where $a_0$ is Bohr radius, $r_s = (3/4 \pi \, n)^{1/3}$, with $n \simeq 2\ 10^{21}$ cm$^{-3}$ and $\rho \simeq$ 150 $\mu\Omega\, cm$\cite{Jovanovic2009b}, we obtain $l \simeq$ 50 \AA.
Thus, the diffusive correction to the ballistic resistance of the contact may be large and contact dependent, due statistical variation of the defects location. This conclusion should however be mitigated: each contact may indeed be made of several contacts in parallel (see below), with each a higher resistance than the measured one. There are indeed examples in the literature for point-contacts well beyond the ballistic regime, that could be adequately fitted within the BTK theory (see e.g. Ref.~\onlinecite{Plecenik1994}, where contact radius could be as large as seven times the mean free path).

The diffusive contribution to the contact resistance may be accounted for by Wexler's formula\cite{Wexler1966}:

\begin{equation}
R = R_{sh} + \beta \, \rho / d
\label{wexler}
\end{equation}

where $\beta \sim 1$ at $l \ll d$. While, in the superconducting regime, the ballistic contribution of the resistance to the spectrum is described by the BTK model, what should be the one for the diffusive part ? Essentially, we expect the diffusive contribution in the superconducting regime to be governed by the critical current density. It may then be easily found that such a critical current is reached in the contact for voltage $eV \sim \Delta d/l$, at which we expect a steep resistance rise, followed by a thermal regime where the resistance grows smoothly. Thus, such a mechanism is able to account for non-spectral features at $eV \gtrsim \Delta$. When $d \sim l$, the presence of a few scattering defects inside the contact more likely yields some mesoscopic fingerprint with applied voltage.

\begin{figure}
\includegraphics[width= \columnwidth]{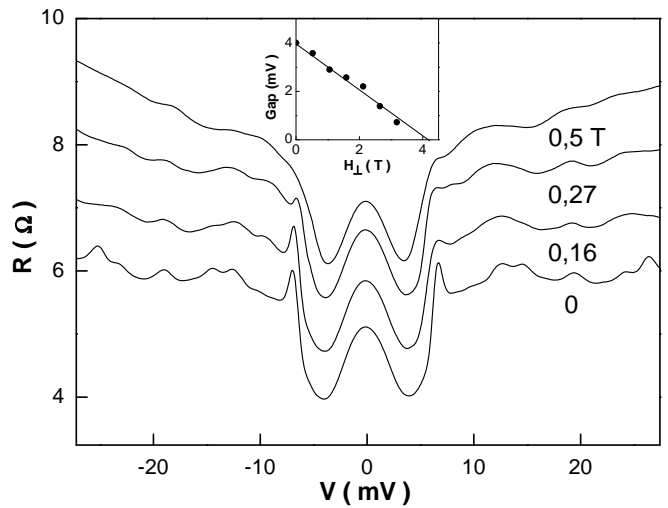}
\caption{Vanishing with magnetic field (transverse magnitude displayed) of both the peaks at the onset of the double minimum structure, and the fingerprints outside of the gap ($T$ = 2 K). Curves have been shifted for clarity. The inset displays the gap from the minimum in $R(V)$.}\label{field}
\end{figure}

For the contact in Fig.~\ref{temp}, we do find evidence for such fingerprints, outside of the voltage range for the double minimum feature, but none within the assumed gap (Fig.~\ref{field}). Similarly, there is no such fingerprints in the data in Fig.~\ref{temp2} either, that would superimpose to one or the other double minimum structures. This suggests that none of the two features in Fig.~\ref{temp2} has a non-spectral origin, and cannot originate either from the addition of independent contacts conductances, as one would expect the fingerprints of the contact with the lower gap to appear within the larger gap. The latter conclusion is also reinforced by the observation that the two contributions, yielding much different characteristic energies, appear to vanish at similar temperatures (Fig.~\ref{gap2}), and that attempts to fit the central feature using two gaps and two independent barrier strength invariably yielded the same barrier parameter for both.

The idea that a critical current may be at the origin of such peaks or shoulders was exploited in Ref.~\onlinecite{Naidyuk1991} and \onlinecite{Haussler1996}. In addition to an anomaly due to the depairing current, a second, sharper anomaly at the edge of a double minimum structure for low impedance contacts was attributed to the resistive state, occurring when the induction at the periphery of the contact exceeds $B_{c1}$. Similar features were also seen for metallic contacts on Niobium, the peak occurring at voltage close to or well above the double minimum, the latter being unambiguously identified as the superconducting gap\cite{Xiong1993}. The critical current model predicts, as was well verified in Ref.~\onlinecite{Haussler1996}:

\begin{equation}
I_c R_N = \frac{\pi \rho B_{c1}}{2 \mu_0}
\label{silsbee}
\end{equation}

using $\rho$ = 150 $\mu \Omega cm$, $\lambda$ = 100 nm\cite{Shengelaya2005,Khasanov2008} and $\xi$ = 50 \AA \cite{Jovanovic2009b} yields $B_{c1} \simeq$ 30 mT and $I_c R_N \simeq$ 60 mV. This evaluation appears too large to account for the data in Fig.~\ref{temp},~\ref{temp2}, but the material parameters could vary greatly in the contact from their bulk value. 

\begin{figure}
\includegraphics[width= \columnwidth]{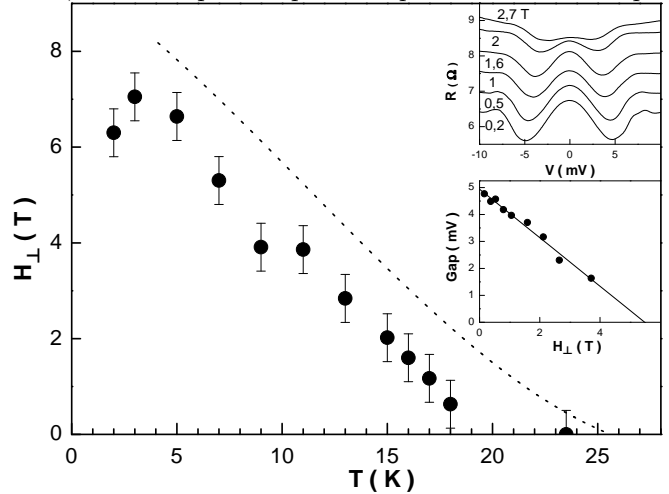}
\caption{Perpendicular component of the magnetic field closing the gap, as obtained from the minimum in $R(V)$. The dotted line is the irreversibility onset, as obtained from resistivity in Ref.~\onlinecite{Jovanovic2009}. Upper inset: spectra for various perpendicular magnetic field component. Lower inset: gap value, as obtained from the minimum in $R(V)$ - the line is a linear fit, extrapolating to zero at the displayed critical field ($T$ = 7 K).}\label{irreversibility}
\end{figure}

However, the observation of sharp peaks away from the shoulders (see e.g. Fig.~\ref{misc}b) suggests that the latter are not due to some critical current effect. Also, it was reported that a magnetic field has no significant effect on the I-V curves, although well above $B_{c1}$\cite{Haussler1996,Xiong1993}, which may be understood as a strong pinning of the regular vortex lattice induced by the applied field\cite{pcs2005}. In Fig.~\ref{field}, it appears that a magnetic field washes out the peaks well before the closing of the double minimum structure (it does so for the peaks in Fig.~\ref{temp2} also). This closing is observed to be linear in the applied magnetic field (assuming quasi two-dimensional behavior\cite{Jovanovic2010}, we only consider the component of the magnetic field perpendicular to the film), as already reported in Ref.~\onlinecite{Naidyuk1998} (Fig.~\ref{irreversibility}, inset). This allows to define a critical field at which the gap extrapolates to zero. As may be seen in Fig.~\ref{irreversibility}, the critical field is close to the irreversibility field obtained from resistivity measurements\cite{Jovanovic2009} (with some temperature shift, which may due to different critical temperatures for the two samples). Thus, this reinforces the idea that this critical field is determined by the vortex lattice melting or depinning\cite{Huxley1993}, whereas the peaks at the edge of the resistance drops have a different origin.

\begin{figure}
\includegraphics[width= \columnwidth]{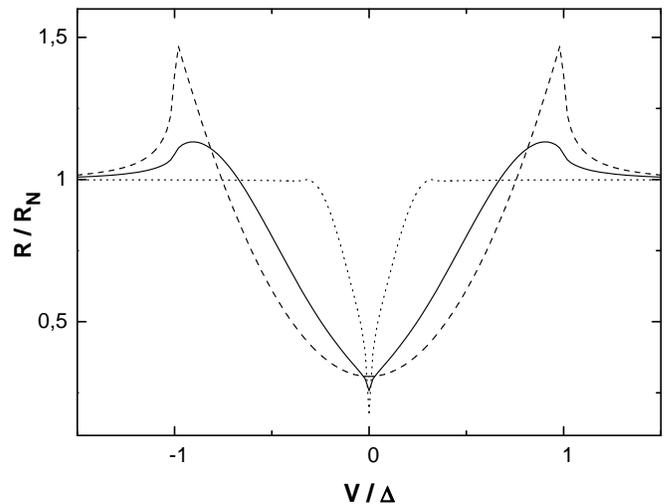}
\caption{Theoretical expectations for a point contact resistance on a $d$-wave superconductor along a (110) interface ($Z=1.1$)\cite{Tanaka1995}. The full line is for a full cylindrical Fermi surface. The dashed line is for antinodal quasiparticles; the dotted one for nodal ones.}\label{theos}
\end{figure}

The washing out of these peaks with a magnetic field well below that for vortex mobility rather suggests an effect related to the phase of the order parameter. It is then appropriate to mention a possible specific contribution of an unconventional gap symmetry to the point-contact spectrum. For a d-wave order parameter, as was recently found in Ref.~\onlinecite{Tomaschko2011} for this compound, one may indeed expect, for some crystallographic orientations of the contact, peaks at the onset of the superconducting gap\cite{Tanaka1995,Kashiwaya1996}. We have plotted in Fig.~\ref{theos} a simulation of the contact resistance for a (110) interface, as obtained from the results in Ref.~\onlinecite{Tanaka1995}, where we have distinguished the contributions from nodal and antinodal quasi-particles, owing to the fact that this e-doped material likely possess both a nodal hole pocket and an antinodal electron one. As may be seen, the contribution of the antinodal particles exhibits peaks at the onset of the superconducting gap. Also, the nodal ones yield a sharp peak at zero bias, which may be compared with the data in Fig.~\ref{misc}a,e. It should also be noticed that, unlike for the conventional gap symmetry, a ratio $R_N/R(0)$ larger than 2 may be expected for the d-wave symmetry. In Fig.~\ref{theos} the ratio for the antinodal particles is $R_N/R(0) \approx$ 3, as is observed in Fig.~\ref{misc}a,c.

So, we do not find a mechanism that would allow us to point unambiguously the non-spectral origin of one or the other characteristic energy. As stated above, the lower energy is in line with previous evaluations of a superconducting gap from tunneling experiments and penetration depth measurements\cite{Chen2002,Khasanov2008}. We could only track the higher characteristic energy down to low temperature for contacts with a low transition temperature, such as in Fig.~\ref{temp2}. This could be related to some doping effect. Indeed, the less doped samples incorporate some excess apical oxygen and, for this compact crystallographic structure, this is susceptible to induce strong electronic inhomogeneity in the CuO$_2$ plane. Considering the potential complexity of the Fermi surface for this electron-doped material, we believe that investigations on better defined crystallographic interfaces should be able to bring new data allowing to resolve this issue.

\newpage

\end{document}